\documentclass{ifacconf}

\usepackage{graphicx}      
\usepackage{natbib}        
\usepackage{color}
\usepackage{amsmath} 
\usepackage{amssymb}  
\usepackage{graphicx}
\usepackage{verbatim}
\usepackage{subfigure}
\usepackage{booktabs}

\newcommand\real{\ensuremath{{\mathbb R}}}
\newcommand\realn{\ensuremath{{\mathbb{R}^n}}}
\newcommand\realnn{\ensuremath{{\mathbb{R}^{n\times n}}}}
\newcommand\mymatrix[2]{\left[\begin{array}{#1} #2 \end{array}\right]}
\newcommand{\smallmat}[1]{\left[ \begin{smallmatrix}#1 \end{smallmatrix} \right]}

\newtheorem{defi}{Definition}
\newenvironment{definition}{\begin{defi}\rm }{\hfill \hspace*{1pt} \hfill $\lrcorner$ \end{defi}}

\begin{document}
\begin{frontmatter}

\title{Antithetic integral feedback for the robust control of monostable and oscillatory biomolecular circuits} 

\thanks[footnoteinfo]{This work is supported in part by DARPA Grant No. HR0011-16-2-0049.} 

\author[NO]{Noah Olsman} 
\author[FF]{Fulvio Forni} 

\address[NO]{Department of Systems Biology, Harvard Medical School, \\ Boston, MA 02215, USA (noah\_olsman@hms.harvard.edu)}
\address[FF]{Department of Engineering, University of Cambridge, \\ Cambridge  CB2 1PZ, UK (f.forni@eng.cam.ac.uk)}

\begin{abstract}                
Biomolecular feedback systems are now a central application area of interest within control theory. While classical control techniques provide invaluable insight into the function and design of both natural and synthetic biomolecular systems, there are certain aspects of biological control that have proven difficult to analyze with traditional methods. To this end, we describe here how the recently developed tools of dominance analysis can be used to gain insight into the nonlinear behavior of the antithetic integral feedback circuit, a recently discovered control architecture which implements integral control of arbitrary biomolecular processes using a simple feedback mechanism. We show that dominance theory can predict both monostability and periodic oscillations in the circuit, depending on the corresponding parameters and architecture. We then use the theory to characterize the robustness of the asymptotic behavior of the circuit in a nonlinear setting.
\end{abstract}


\end{frontmatter}

\section{Introduction}
\label{sec:introduction}


Feedback regulation is ubiquitous in biology, playing a crucial role in high-level phenomena, such as sensory perception in animals (\cite{wiener1948cybernetics, nakahira2019diversity}), all the way down to the most basic processes in life, like the regulation of amino acid biosynthesis (\cite{monod1971chance}). As we refined our understanding of molecular biology, it became abundantly clear that life not only relies heavily of feedback control, but that this control is often extremely precise and robust (\cite{barkai1997robustness}). It has become a central goal of biological engineering to implement synthetic feedback controllers in cellular systems, the performance of which we hope will rival that found in nature (\cite{del2016control}).

In the past few years, a great deal of progress has been made towards this goal of designing universal feedback controllers that can easily be used in a wide variety of contexts (\cite{aoki2019universal,samaniego2017ultrasensitive,chevalier2019design,becskei2000engineering, huang2018quasi}). Among these, a particularly promising architecture relies on what has been named the Antithetic Integral Feedback (AIF) circuit, which has been shown to implement integral feedback using a simple irreversible bimolecular interaction as the primary feedback mechanism (\cite{briat2016antithetic}). This architecture has several appealing properties: it relies on a single molecular interaction that appears in a variety of natural contexts (e.g., the nearly irreversible sequestration of sigma factor/antisigma factor pairs), and the model of its dynamics is simple enough that it is possible to prove general results about stability and optimality (\cite{aoki2019universal,olsman2019hard}). Recent theoretical work has focused both on global results for arbitrary plant-controller systems and local results focused on applying classic tools from control theory to simplified models of closed-loop dynamics (\cite{qian2018multi,qian2018realizing, olsman2019hard, olsman2019architectural, briat2016antithetic,briat2018antithetic}).

While much progress has been made, there are still some basic properties of the system that have, so far, proven difficult to address theoretically. For example, it was shown via simulation that, for some plant models, local instability of the closed-loop system yields stable limit cycles globally (\cite{briat2016antithetic}). While we have a fairly good understanding of local stability for some biologically realistic parameter regimes, there is still relatively little that is understood about the specifics of the AIF circuit's global, nonlinear behavior. This is in part due to the fact that the simplest models of the system that can produce limit cycles have at least four states (\cite{olsman2019hard}), making it difficult to use any classical results from the theory of planar systems (e.g., the Poincar\'e-Bendixson theorem). It has been similarly difficult to construct a global Lyapunov function that would facilitate the use of other nonlinear tools, such as LaSalle's Invariance Principle or a generalized Hopf bifurcation analysis. 

To address these issues, we will use the tools of dominance theory to study the AIF circuit. By adopting a differential perspective, dominance theory generalizes classical tools from linear system theory to the nonlinear setting, for the analysis of multistable and oscillatory closed-loop systems. In this paper we will illustrate how to use dominance theory to derive novel results about a biological system that, so far, has proven difficult to analyze with classical tools. We show how the AIF system achieves closed-loop regulation  and how the same device can be used to design robust oscillatory circuits. The analysis emulates the approach pursued for mechanical and electrical system (\cite{Forni_Sepulchre_tutorial_cdc2014, Miranda-Villatoro2018b}), demonstrating the potential dominance theory has for robustness analysis
of AIF circuits.

\section{Antithetic integral feedback circuits}
\label{sec:antithetic_integral_feedback_circuit}

The AIF controller, which we denote $\Sigma_z$, follows the dynamics:
\begin{equation}
\Sigma_z:
\left\{
\begin{split}
\dot{z}_1 & = \mu - \eta z_1z_2, \\
\dot{z}_2 & = u_c - \eta z_1 z_2, \\
y_c &= z_1, 
\end{split} 
\right.
\label{eq:sequestration}
\end{equation}
where $z_1 \geq 0$ can be thought of as the actuator species and $z_2 \geq 0$ the sensor species. $\mu$ is the reference input (for adaptation).
$\eta$ is the rate at which $z_1$ and $z_2$ bind and are jointly removed from the system. $\eta$ typically captures a sequestration rate, but can also represent any interaction in which two species are mutually inactivated. The key property here is that $\eta$ is exactly identical for both $z_1$ and $z_2$. 

Neither $z_1$ nor $z_2$ in system \eqref{eq:sequestration} directly implements an integrator. Their difference $z = z_1 - z_2$, however, follows the dynamics 
\begin{equation}\label{eq:zdiff}
    \dot{z} = \dot{z}_1 - \dot{z}_2 = \mu - u_c \implies z = \int_0^t \mu - u_c dt,
\end{equation}
which encodes a virtual integrator internal to the system's state. 
At equilibrium, we have that  
\[\dot{z} = 0 \implies u_c = \mu.\]
This means that, for any plant $\mathcal{P}$ with input $u$ and output $y$, 
any stable closed-loop interconnection given by 
\begin{subequations}
\label{eq:closed_loop_interconnection}
\begin{flalign}
u & = y_c \label{eq:u=y_c}\\
u_c & = y \label{eq:u_c=y}
\end{flalign}
\end{subequations}
must asymptotically yield $y = \mu$, demonstrating the robustness of the closed-loop equilibrium to parametric variations. 

For the purposes of this article, we will study the Jacobian of system \eqref{eq:sequestration}:
\begin{equation}
\dot{\delta} z = 
\mymatrix{cc}{-\eta z_2 & -\eta z_1 \\ -\eta z_2 & -\eta z_1} \delta z + 
\mymatrix{c}{0 \\ 1} \delta u_c  \ .
\label{eq:dsequestration}
\end{equation}
A crucial observation is that system \eqref{eq:dsequestration} is not the linearization of the system around a specific equilibrium. Rather, it is the linearization of the system along any possible trajectory $z(\cdot)$ of system \eqref{eq:sequestration}. The analysis in the next sections depends crucially on this general parameterization of the Jacobian (i.e., a differential perspective of dynamics).

\section{Dominance theory}
\label{sec:dominance}
\subsection{The workflow of dominance analysis}
Dominance theory provides a set of tools to study nonlinear models which have behavior that is not constrained to a single stable equilibrium. The theory takes nonlinear models with parameters in a certain range and characterizes their behavior through Lyapunov-like linear matrix inequalities (LMIs).
Dominance theory was introduced in \cite{Forni2019,Miranda-Villatoro2018} to study systems that have several equilibria or that oscillate. The intuition for the theoretical framework is that the dynamics of a $p$-dominant system 
\begin{equation}
    \dot{\xi} = f(\xi) \qquad \xi \in \real^n 
    \label{eq:sys_dom}
\end{equation} 
can be split into fading sub-dynamics of dimension $n-p$ and dominant sub-dynamics of dimension $p$, constraining the system's steady-state behavior. Its attractors must thus be compatible with a system of dimension $p$. For $1$-dominant systems this means that every bounded trajectory converges to an equilibrium. For $2$-dominant systems, bounded trajectories converge to a simple attractor, compatible with a planar dynamics.

Dominance is studied differentially, by looking at the linearized dynamics along system trajectories, namely
\begin{equation}
    \left\{ 
    \begin{array}{rcl}
    \dot{\xi} &=& f(\xi) \\
    \dot{\delta} \xi &=& \partial f(\xi) \delta \xi 
    \end{array}
    \right.
    \qquad (\xi,\delta \xi) \in \realn \times \realn \ ,
    \label{eq:dsys_dom}
\end{equation}
where $\partial f(\xi)$ is the Jacobian of $f$ at $\xi$\footnote{For simplicity, all functions in the paper are differentiable.}. 
\begin{definition}
\label{def:dominance}
The nonlinear system \eqref{eq:sys_dom} is $p$-dominant with rate $\lambda \geq 0$ if there exist a symmetric matrix $P$ with inertia $(p, 0, n-p)$ and a positive constant $\varepsilon$ such that
\begin{equation}
\label{eq:dominance}
\mymatrix{c}{\dot {\delta \xi} \\ \delta \xi}^T \!
\mymatrix{cc}{
0 & P \\ P & 2\lambda P + \varepsilon I
}
\mymatrix{c}{\dot {\delta \xi} \\ \delta \xi}
\leq 0
\end{equation}
along the trajectories of \eqref{eq:dsys_dom}.
\end{definition}
We recall that a matrix $P\in\realnn$ with inertia $(p, 0, n-p)$ has $p$ negative eigenvalues and $n-p$ positive eigenvalues. \eqref{eq:dominance} is equivalent to finding a uniform solution $P$ to the matrix inequality
\begin{equation}
    \partial f(\xi)^T P + P \partial f(\xi) + 2 \lambda P \leq - \varepsilon I \qquad \forall \xi \in \realn
\label{eq:dominance_LMI}
\end{equation}
(for some $\varepsilon > 0$). 
Inequality \eqref{eq:dominance_LMI} is a standard Lyapunov inequality adapted to the linearization, where positivity of $P$ has been replaced by a constraint on the matrix inertia. The constraint on the inertia is used to separate dominant and fading dynamics. In fact, for linear systems $f(\xi) = A\xi$, LMI \eqref{eq:dominance_LMI} implies that $A$ has $p$ dominant eigenvalues to the right of  $-\lambda$, and $n-p$ stable eigenvalues to the left of $-\lambda$, \cite[Proposition 1]{Forni2017a}.

The use of dominance theory in nonlinear control is motivated by the following theorem
\begin{thm} {\cite[Corollary 1]{Forni2019}}. \\
\label{thm:behavior}
Every bounded solution of a $p$-dominant system with rate $\lambda \geq 0$ asymptotically converges to
\begin{itemize}
    \item a unique fixed point if $p=0$;
    \item a fixed point if $p=1$;
    \item a simple attractor if $p = 2$, that is, a fixed point, a set of
fixed points and connecting arcs, or a limit cycle.
\end{itemize}
\end{thm}

\begin{figure}[htbp]
\begin{center}
\includegraphics[width=\columnwidth]{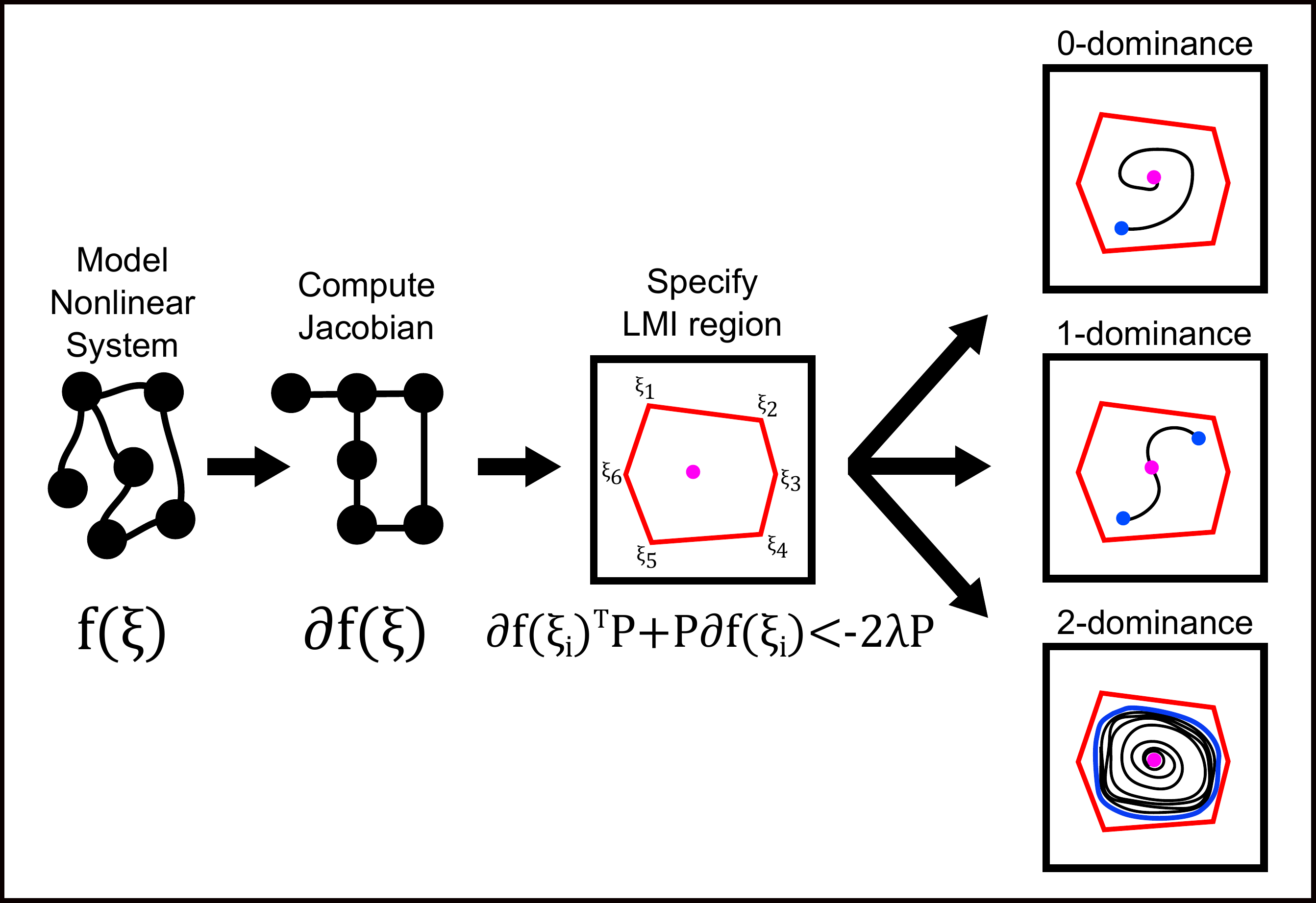} \label{fig:cartoon}
\caption{\small\textbf{The workflow of dominance theory:} 
1) start with a nonlinear model of interest; 
2) compute the corresponding Jacobian; 
3) specify points $\xi_i$, the convex hull of which define a region of interests (red);
finally 4) use LMIs to generate a certificate of dominance that describes the system's limiting behavior within the region (blue). 
In the last panel, we see that 0-dominance corresponds to single equilibrium, 1-dominance corresponds to multiple equilibria, and 2-dominance corresponds to a simple attractor.}
\end{center}
\end{figure}

It is important to note that LMI \eqref{eq:dominance_LMI} is numerically tractable. It can be reduced to a finite set of inequalities built from the linearization of the system computed at a set of points forming a convex hull of the region of interest (both in parameter space and state space). The whole workflow is illustrated in Figure \ref{fig:cartoon}.

\subsection{Dominance design through necessary conditions}
\label{sec:dominance_necessary_conditions}
The eigenvalues of the Jacobian $\partial f(x)$ of a $p$-dominant system are always split into two groups, with $p$ eigenvalues to the right of $-\lambda$ and the remaining ones to the left (at each $x$). This necessary condition opens the way to the use of root locus methods and Nyquist criterion in nonlinear control.

Consider the (open) nonlinear system 
\begin{equation}
    \dot{\xi} = f(\xi) + Bu \,,\ y = c(\xi) \qquad \xi \in \real^n, y \in \real, u \in \real
    \label{eq:osys_dom}
\end{equation} 
and the closed-loop system arising from \eqref{eq:osys_dom} through the feedback interconnection 
\begin{equation}
    u = -ky \,  .
    \label{eq:feedback}
\end{equation}
If the closed-loop system \eqref{eq:osys_dom},\eqref{eq:feedback} is $p$-dominant with rate $\lambda$ then the closed-loop Jacobian
\begin{equation}
\label{eq:closed_loop_jacobian}
A_{cl}(\xi) := \partial f(\xi) +  k B\partial c(\xi)
\end{equation}
satisfies \eqref{eq:dominance_LMI} and the splitting of its eigenvalues can be predicted through root locus analysis and Nyquist criterion applied to the (frozen) state-dependent transfer function
\begin{equation}
    G(\xi,s) := \partial c(\xi)(sI - \partial f(\xi))^{-1}B \,.
    \label{eq:Gx}
\end{equation}

\begin{thm}[Root locus and Nyquist criterion] 
\label{thm:rootlocus_nyquist}
$ $\\
Suppose that for some $k = k^*$ the closed-loop system described by \eqref{eq:osys_dom},\eqref{eq:feedback} is $p$-dominant with rate $\lambda$. Suppose that the pairs $(\partial f(\xi),B)$ and $(\partial f(\xi),\partial c(\xi))$ are controllable and observable at $\xi$, respectively.
Then, 
\begin{itemize}
\item 
the root locus of $G(\xi,s)$ for the feedback gain $k^*$ has $p$ poles whose real part is greater than $-\lambda$ and $n-p$ poles whose real part is smaller than $\lambda$, for each $\xi\in \realn$;
\item
the Nyquist plot of 
$G(\xi,-\lambda + j\omega)$ for $\omega \in \real$
encircles the point the point 
$-\frac{1}{k^*}$ exactly $(p - q_\xi)$ times in the clockwise direction, where $q_\xi$ is the number of poles of $G(\xi,s)$ whose real part is greater than $-\lambda$.
\end{itemize}
\end{thm}
\begin{pf}
Under controllability and observability assumptions, the first item follows directly from the fact that the roots of $1+k G(\xi,s)$ correspond to the eigenvalues of $A_{cl}(\xi)$. In a similar way, the second item follows from the argument of the proof of \cite[Theorem 3.1]{Miranda-Villatoro2018}.
\hfill $\square$
\end{pf}

In what follows we will use the necessary conditions of Theorem \ref{thm:rootlocus_nyquist} as guidelines for control design, since they define minimal requirements that a dominant closed-loop system has to satisfy. Root locus and Nyquist diagrams will be used to find model parameters that guarantee a suitable splitting of the closed-loop eigenvalues, taking into account classical robustness considerations. Validation will then be certified through linear matrix inequalities \eqref{eq:dominance_LMI} applied to the closed-loop Jacobian in equation \eqref{eq:closed_loop_jacobian}.

\section{First-order production system: homeostasis and oscillations}

\subsection{Linear feedback}\label{sec:fop}

We will start by analyzing the linear plant model
\begin{equation}
\Sigma_x:
\left\{
\begin{split}
    \dot{x_1} &= \theta_1 u - \gamma x_1, \\
    \dot{x_2} &= k x_1 - \gamma x_2 \\
    y &= x_2
\end{split}
\right.
\label{eq:cl1}
\end{equation}
We refer to \eqref{eq:cl1} as \textit{first-order production} system. Here $\theta_1$ and $k$ represent first-order production rates for each species, and $\gamma$ represents a common first-order degradation rate. It has been demonstrated via simulation that this simple model in closed loop with the nonlinear AIF controller can exhibit stable limit cycles (\cite{briat2016antithetic}), which appear to arise when the linearized closed-loop system becomes locally unstable. In the limit of large $\eta$ this corresponds to the parametric condition
$ \sqrt[3]{\frac{\theta_1\theta_2 k}{2}} < \gamma$,
as shown in (\cite{olsman2019hard}). 
We will then extend our study to a plant with the same qualitative architecture but where the interconnections are modeled with more realistic saturation behavior, such as with the Hill function $\theta_1(u) = u^N/(k_1 + k_2 u^N)$.

\begin{figure}[htbp]
\begin{center}
\includegraphics[width=0.8\columnwidth]{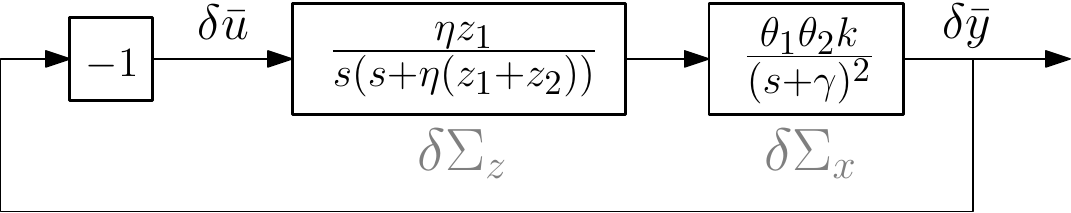}  
\caption{\small
Linearized closed-loop system diagram of
\eqref{eq:sequestration}, \eqref{eq:u=y_c}, \eqref{eq:cl1} with $\bar{u} = -\bar{y}$.
}
\label{fig:cl1}
\end{center}
\end{figure}

We represent the closed loop of AIF circuit and first-order production system 
\eqref{eq:sequestration}, \eqref{eq:closed_loop_interconnection}, \eqref{eq:cl1}
as the interconnection of
\eqref{eq:sequestration}, \eqref{eq:u=y_c}, \eqref{eq:cl1}
with input $\bar{u} = u_c$ and output $\bar{y} = -y$, through negative feedback $\bar{u} = -\bar{y}$. The linearized closed loop is shown in Figure \ref{fig:cl1}. The state-dependent linearized transfer function from $\delta \bar u$ to $\delta \bar y$ reads
\begin{equation}
T_{\delta \bar{u} \to \delta \bar{y}}(z,s) = \frac{\theta_1 \theta_2 \eta k z_1}{ s(\gamma + s)^2 (s + \eta (z_1+z_2))} \ .
\end{equation}

For any fixed $z$, the root locus of $T_{\delta \bar{u} \to \delta \bar{y}}(z,s)$ shows that 
poles move towards infinity as $\theta_1\theta_2 k$ becomes large, along four asymptotes oriented as $\frac{\pi}{4} + q \frac{\pi}{2}$ 
where $q \in \{0,1,2,3\}$. In agreement with Theorem \ref{thm:rootlocus_nyquist}, this suggests a potential transition from $0$-dominance to $2$-dominance, as $\theta_1\theta_2 k$ increases, with poles separated into stable and dominant pairs, respectively to the left and to the right of the ($z$-dependent) centroid.
Similar observations follow from the Nyquist plot, which makes no rotations around $-1$ for $\theta_1\theta_2 k$ small, but has two clockwise rotations around $-1$ for large $\theta_1\theta_2 k$. 

For simplicity, we develop the details of the analysis for fixed parameters $\mu = 2$, $\eta = 10$ (controller) and $\theta_1 = 1$, $\gamma = 1$ (plant). The analysis can be easily adapted to different parameters.

\begin{figure}[htbp]
\begin{center}
\subfigure[]{\includegraphics[width=0.48\columnwidth]{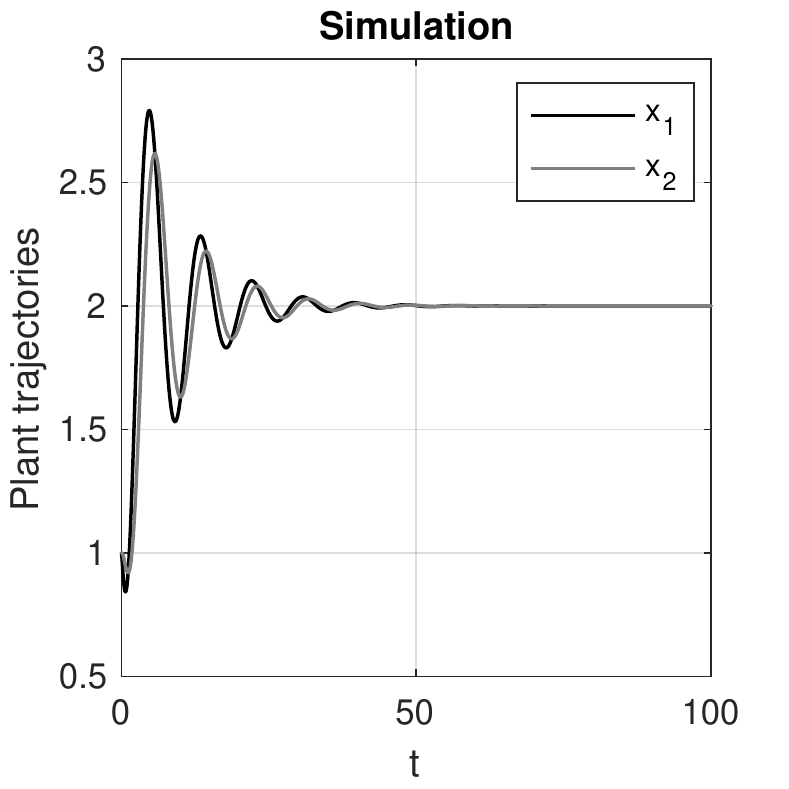} \label{fig:case1_0dom_a}}
\subfigure[]{\includegraphics[width=0.48\columnwidth]{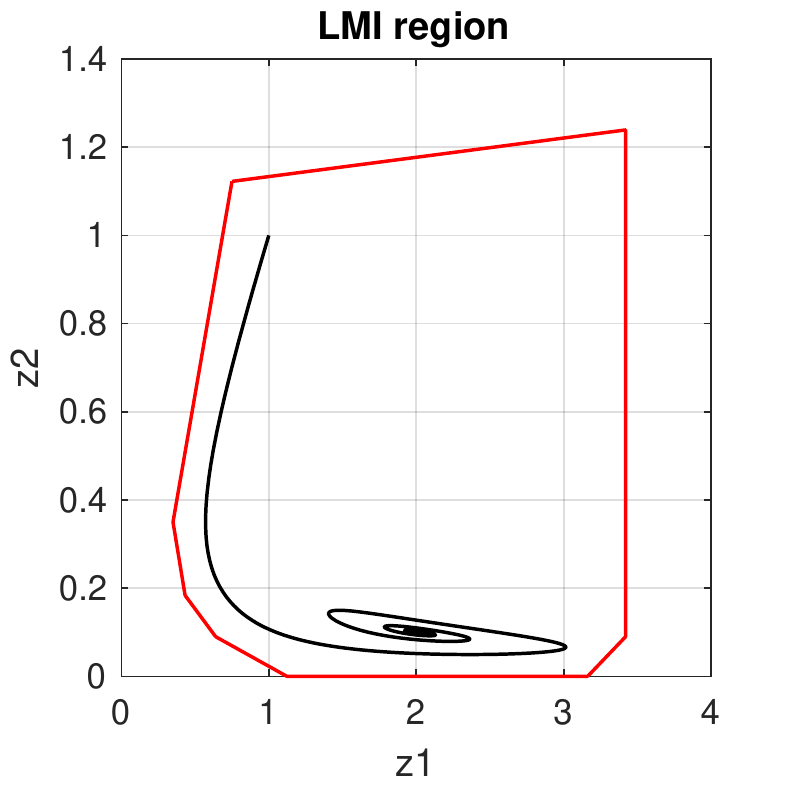}  \label{fig:case1_0dom_b}} 
\subfigure[]{\includegraphics[width=0.48\columnwidth]{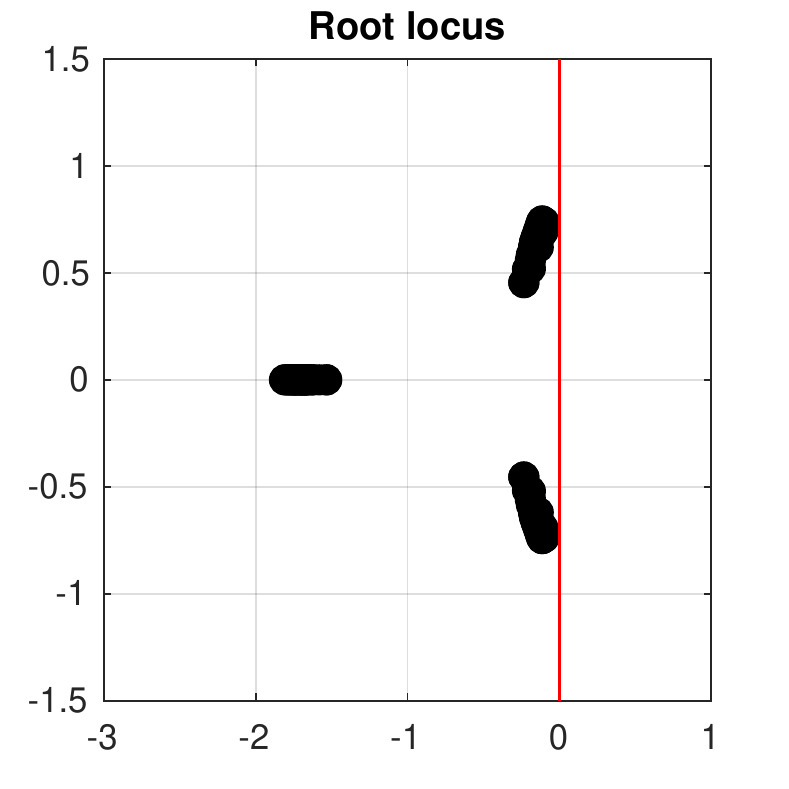} \label{fig:case1_0dom_c}}
\subfigure[]{\includegraphics[width=0.48\columnwidth]{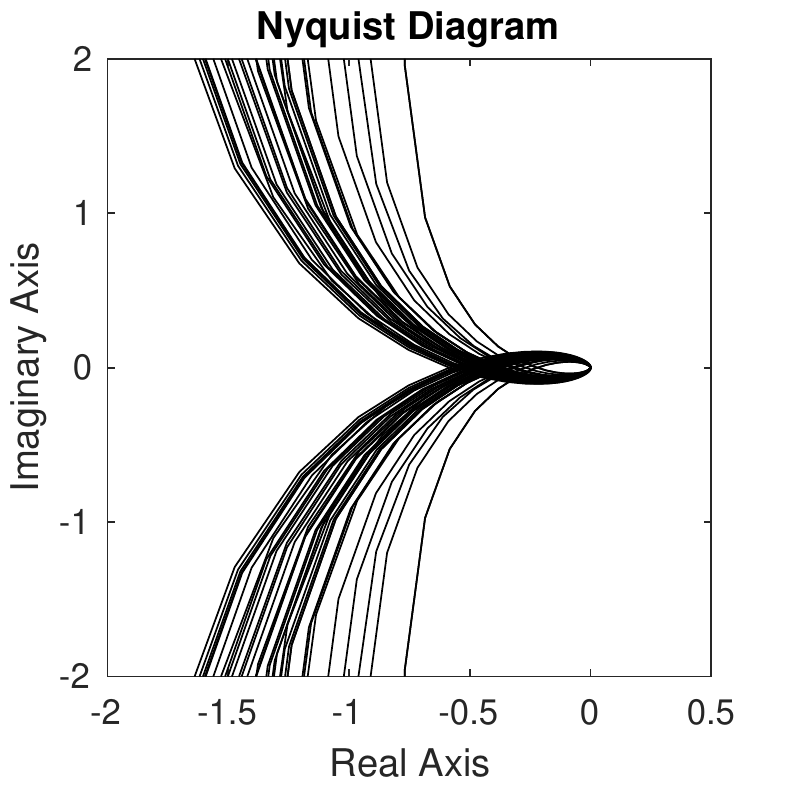} \label{fig:case1_0dom_d}}
    \vspace{-2mm}
    \caption{\small (a,b): trajectories (time-trajectories for $x$, $z$-plane projection) for $\theta_2=k=1$.
    The red curve in (b) delimits the boundary of the convex region $\mathcal{R}_0$.
    (c): closed-loop Jacobian eigenvalues for $(x,z) \in \real^2\times \mathcal{R}_0$. 
    (d): Nyquist locus of 
    $T_{\delta \bar u \to \delta \bar y}(z,j\omega)$ for $\omega \in \real$ and $z \in \mathcal{R}_0$.}
    \label{fig:case1_0dom} 
    \end{center} 
\end{figure}

As a first case, consider $\theta_1\theta_2 k = 1$  (small). The Nyquist plots in Figure \ref{fig:case1_0dom_d} samples $T_{\delta \bar{u} \to \delta \bar{y}}$ for $z$ constrained to the convex red region $\mathcal{R}_0$ in Figure \ref{fig:case1_0dom_b}. Indeed, accordingly to Theorem \ref{thm:rootlocus_nyquist}, the Nyquist plots are all compatible with $0$-dominance. Compatibility is reinforced by Figure \ref{fig:case1_0dom_c}, which shows that the closed-loop Jacobian eigenvalues have negative real part for all $x\in \real^2$ and all $z\in \mathcal{R}_0$. From the perspective of the LMI \eqref{eq:dominance_LMI}, the closed loop is $0$-dominant with rate $\lambda = 0$ for $x\in \real^2$ and $z\in \mathcal{R}_0$. This is certified by the first positive-definite matrix $P$ in Table \ref{table:p}, which is a solution to \eqref{eq:dominance_LMI} for $\xi = [z^T,x^T]^T$ and for  $f$ representing the closed-loop dynamics. The solution is computed by convex relaxation, taking advantage of the finite number of vertices of $\mathcal{R}_0$.

From Theorem \ref{thm:behavior}, it follows that every attractor in $\mathcal{R}_0\times \real^2$ is necessarily a fixed point, as illustrated by plant time trajectories in Figure  \ref{fig:case1_0dom_a} and controller phase plane trajectories in Figure \ref{fig:case1_0dom_b}. Local asymptotic stability of the equilibrium 
$
\xi_e = [\frac{\mu}{k\theta_1\theta_2} \, \frac{k\theta_1\theta_2}{\eta} \, \frac{\mu}{\theta_2} \, \frac{\mu}{k\theta_2}]^T
= [2 \ 0.1 \ 2 \ 2]^T
$ 
follows from standard Lyapunov argument, using $(\xi-\xi_e)^T P(\xi-\xi_e)$ as a Lyapunov function. We observe that the region admits a unique fixed point, since the presence of two fixed points $\xi_e$ and $\xi_e'$ would mean 
\[\frac{d}{dt} (\xi_e'-\xi_e)^T P(\xi_e'-\xi_e) = 0,\]
which is not compatible with LMI \eqref{eq:dominance_LMI}.

\begin{figure}[t]
\begin{center}
\subfigure[]{\includegraphics[width=0.48\columnwidth]{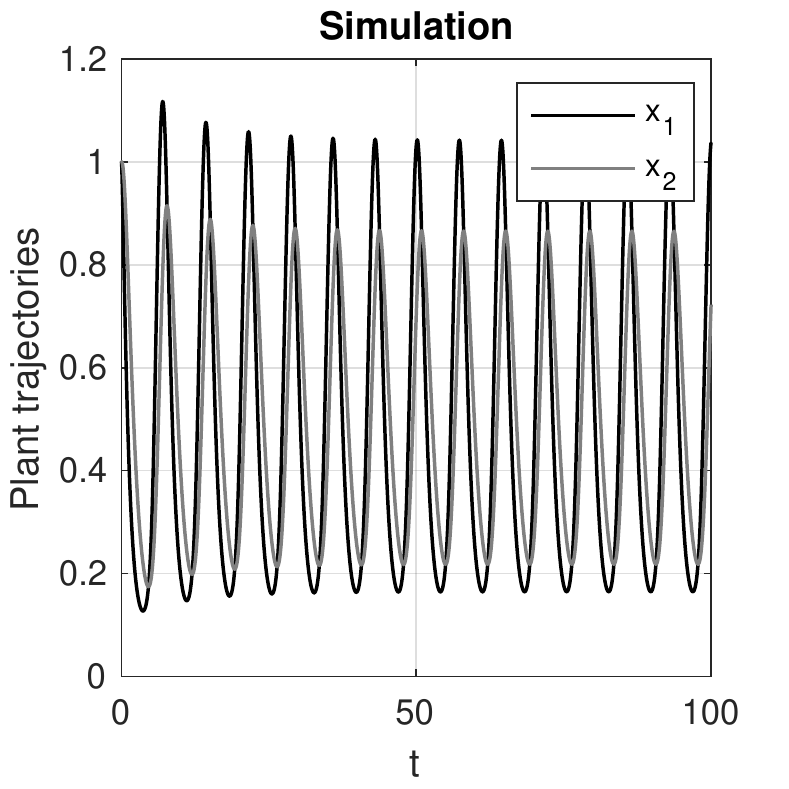} \label{fig:case1_2dom_a}}
\subfigure[]{\includegraphics[width=0.48\columnwidth]{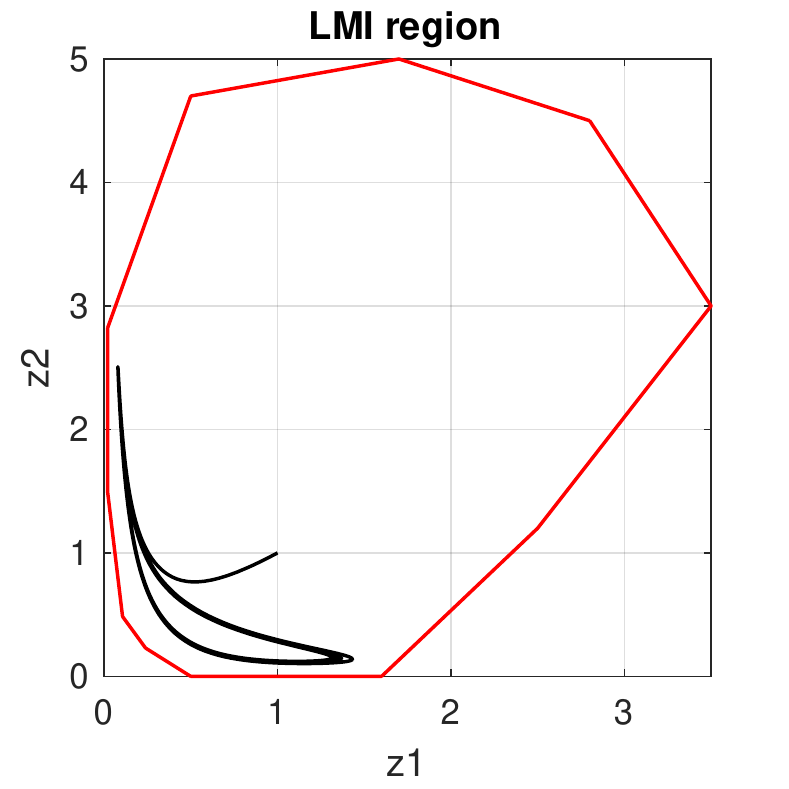} \label{fig:case1_2dom_b}}
\subfigure[]{\includegraphics[width=0.48\columnwidth]{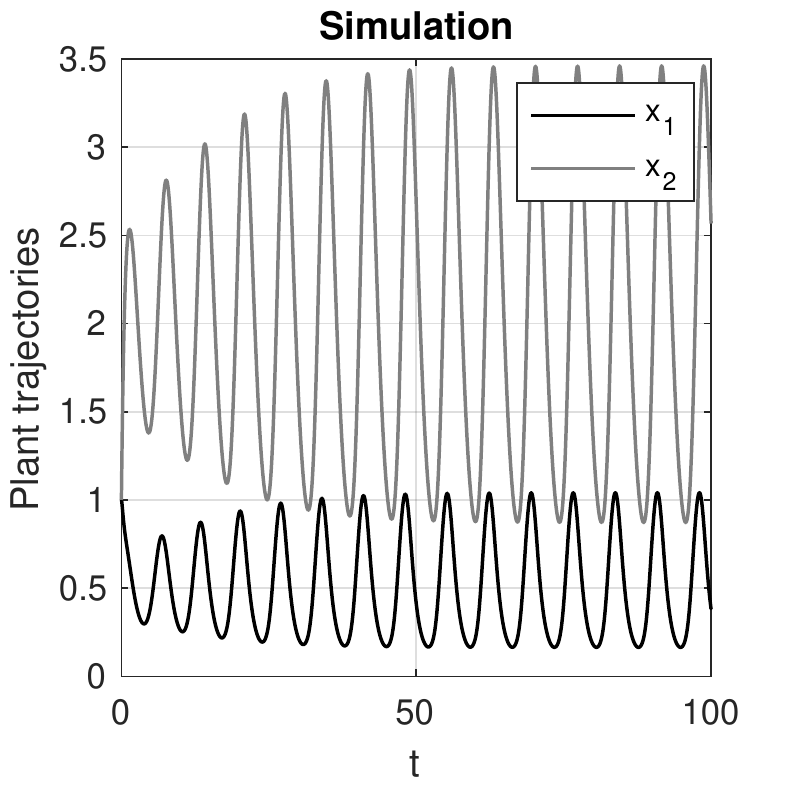} \label{fig:case1_2dom_d}}
\subfigure[]{\includegraphics[width=0.48\columnwidth]{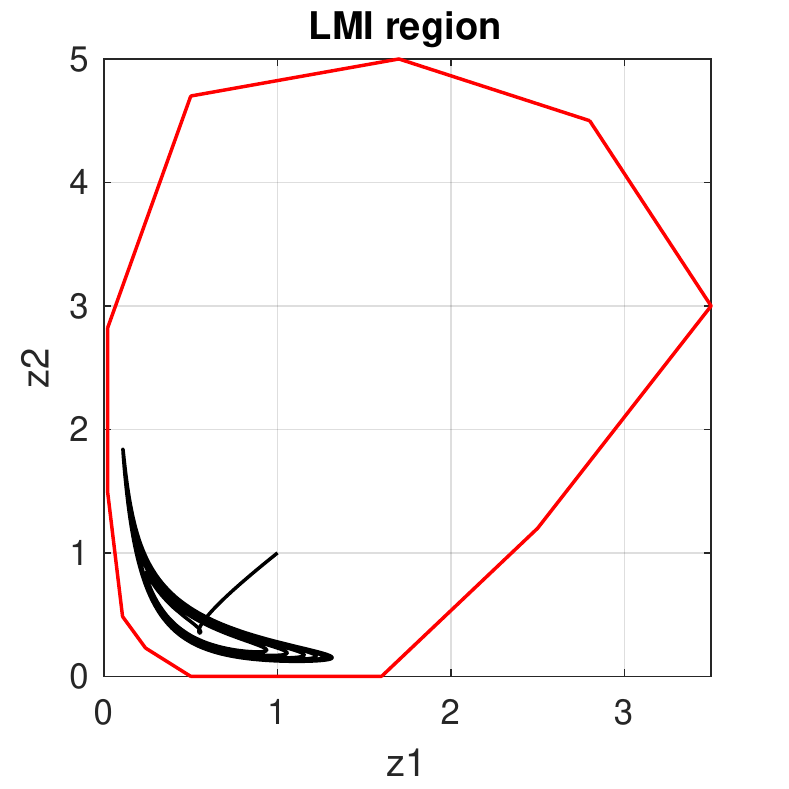} \label{fig:case1_2dom_e}}

\subfigure[]{\includegraphics[width=0.48\columnwidth]{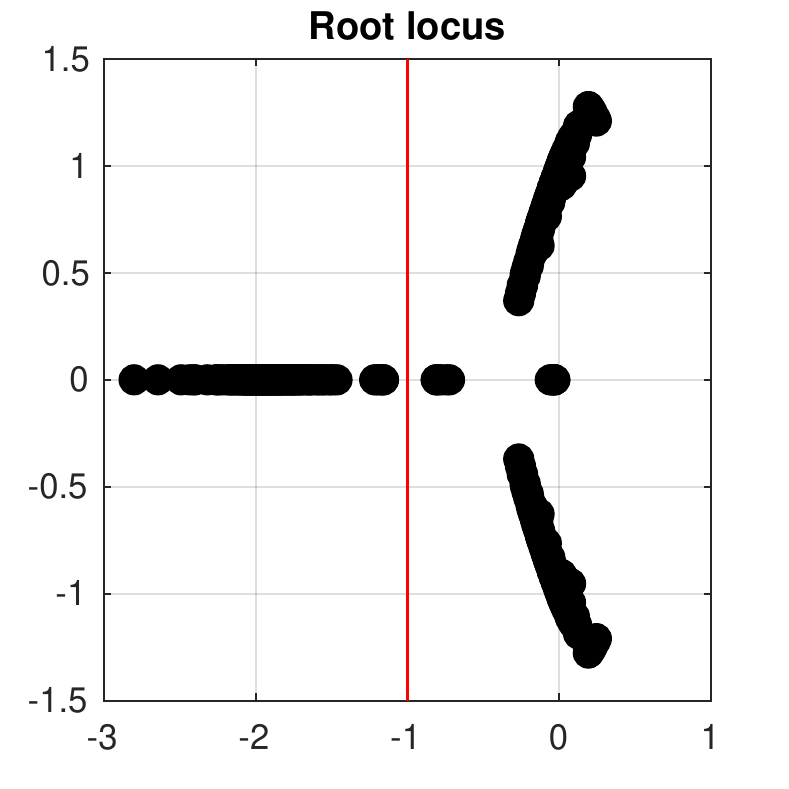} \label{fig:case1_2dom_c}}
\subfigure[]{\includegraphics[width=0.48\columnwidth]{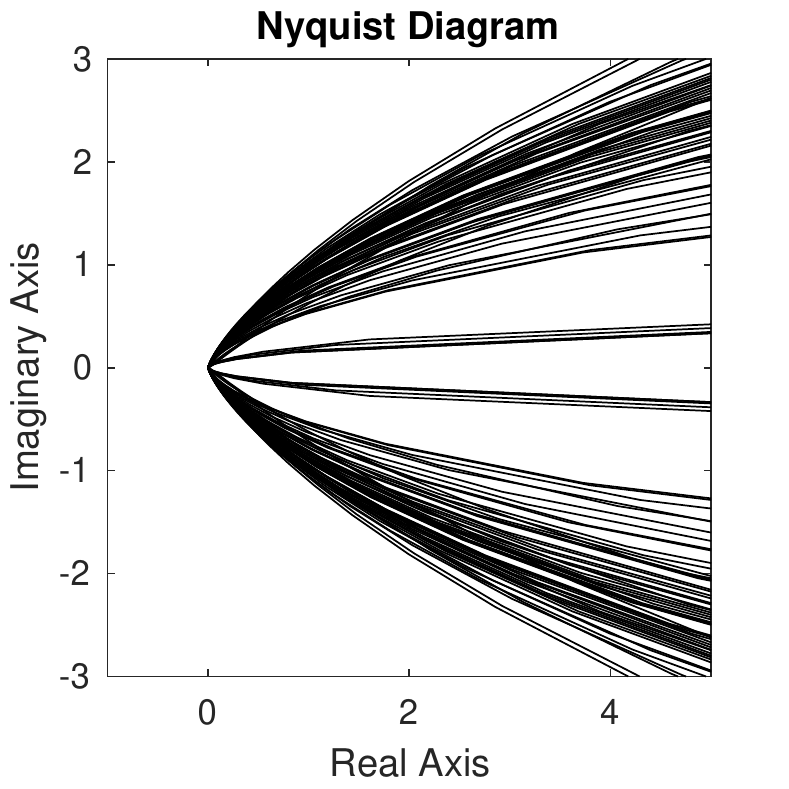} \label{fig:case1_2dom_f}}
    \vspace{-2mm}
    \caption{\small (a,b): trajectories for $k=1$,$\theta_2=4$.
    (c,d): trajectories for $\theta_2=1$,$k=4$. 
    The red curve in (b,d) delimits the convex region $\mathcal{R}_2$.
    (e): closed-loop Jacobian eigenvalues for $x \in \real^2$ and $z\in \mathcal{R}_2$. 
    (f): Nyquist locus of $T_{\delta \bar u \to \delta \bar y}(z,-\lambda + j\omega)$, for $\omega \in \real$,
    $z \in \mathcal{R}_2$, and $\lambda = 1$.}
    \label{fig:case1_2dom} 
\end{center} 
\end{figure}

Figure \ref{fig:case1_0dom_d} makes clear that $0$-dominance is lost
by increasing the feedback gain $\theta_1\theta_2k$.
Consider the convex red region $\mathcal{R}_2$ defined by 
the red curve in Figure \ref{fig:case1_2dom_b} and take
$\theta_1\theta_2k = 4$. For $x\in \real^2$ and $z\in \mathcal{R}_2$, 
Figure \ref{fig:case1_2dom_c} shows that the closed-loop Jacobian eigenvalues 
are not compatible with $0$-dominance but still support $2$-dominance with rate $\lambda = 1$
(from Theorem \ref{thm:rootlocus_nyquist}).
This is reinforced by the Nyquist plots in Figure \ref{fig:case1_2dom_f}, 
which satisfies Theorem \ref{thm:rootlocus_nyquist}.
From  the  perspective  of  the  LMI \eqref{eq:dominance_LMI},
the closed-loop system is $2$-dominant with rate $\lambda = 1$ 
for $x\in \real^2$ and $z\in \mathcal{R}_2$, as certified
by the second and third matrices in Table \ref{table:p}.

From Theorem \ref{thm:behavior}, any attractor in 
$\mathcal{R}_2\times \real^2$ is a simple attractor. 
Furthermore, since the region contains only unstable equilibria
(by computing the closed-loop Jacobian eigenvalues 
for $z = [\frac{\mu}{k\theta_1\theta_2} \, \frac{k\theta_1\theta_2}{\eta}]^T 
= [0.5 \ 0.4]^T$),
the attractor must be a limit cycle, as 
illustrated by Figures 
\ref{fig:case1_2dom_a}-\ref{fig:case1_2dom_e}.

The analysis above shows how the feedback gain (of the linearization) modulates
between monostable and oscillatory behaviors. Gain tuning is based on
intuitive linear feedback considerations, from Nyquist diagrams and root locus argument, 
followed by formal certification through linear matrix inequalities. The analysis
illustrates the versatility of the antithetic integral control, which enables
circuits capable of homeostatic regulation but also of stable oscillations.

\begin{table}[htbp]
\begin{center}
\begin{tabular}{ c c}
\toprule
    $\begin{array}{l}
	\theta_2 =1 \\
	k=1 \\ 
	0\mbox{-dom} 
	\end{array}$ &
	$\smallmat{
   82.7650 & -79.6308 & -24.2526 & -42.5767 \\
  -79.6308 &  89.5784 &  24.7259 &  43.0863 \\
  -24.2526 &  24.7259 &  79.5742 &  24.4801 \\
  -42.5767 &  43.0863 &  24.4801 &  83.9389
	}$ \vspace{1mm}\\
	\midrule
    $\begin{array}{l}
	\theta_2 =4 \\
	k=1 \\ 
	2\mbox{-dom} 
	\end{array}$ &
	$\smallmat{
   -6.0678 &  18.1011 & -60.7826 &  57.3881 \\
   18.1011 & -14.2661 &  60.3814 & -57.2023 \\
  -60.7826 &  60.3814 &   0.7924 & -42.0355 \\
   57.3881 & -57.2023 & -42.0355 & -228.9326
	}$ \vspace{1mm}\\
	\midrule
	$\begin{array}{l}
	\theta_2 =1 \\
	k=4 \\ 
	2\mbox{-dom} 
	\end{array}$ &
	$\smallmat{
  -15.2507 &  32.0157 & -83.0632 &  26.8417 \\
   32.0157 & -26.7504 &  82.5277 & -26.8390 \\
  -83.0632 &  82.5277 &  -4.1886 & -15.9095 \\
   26.8417 & -26.8390 & -15.9095 & -26.8531
	}$ \\ 
\bottomrule 	\addlinespace[1.5ex]
\end{tabular}
	\caption{\small Matrices $P$ for dominance, for different values of parameters $\theta_2$ and $k$.}
	\label{table:p}
\end{center}
\end{table}

\subsection{Nonlinear feedback and robustness}
\label{sec:robustness}

The simplest way to develop robustness analysis is to encode it via linear matrix inequalities, like those already used in LMI \eqref{eq:dominance_LMI}. This can be achieved by considering a perturbed closed-loop Jacobian $\partial f(\xi)+ \Delta(\xi)$ where $\Delta$ captures a set of unknown bounded perturbations $\mathcal{D}$ on the system Jacobian.

The existence of a uniform solution $P$ to 
\begin{equation}
[\partial f(\xi) + \Delta(\xi)]^T P + P [\partial f(\xi) + \Delta(\xi)]+ 2 \lambda P \leq - \varepsilon I
\label{eq:robust_dominance_LMI}
\end{equation}
with fixed inertia for any $\xi$ any $\Delta \in \mathcal{D}$ ($\varepsilon > 0$) guarantees robust dominance. As in LMI \eqref{eq:dominance_LMI}, such a $P$ can be found through convex relaxation, based on the linearization of the system computed at a finite number of points, both in state space and in parameter space. 

In general, several classical tools from robust linear control have been extended to dominant systems, like robustness margins, 
circle criteria, and small gain theorem. The intuition is that Nyquist diagrams and root locus get affected proportionally to the magnitude of the perturbation, therefore small perturbations do not change the dominance of the system (as in classical linear robust analysis). The interested reader is referred to \cite{Padoan2019b,Padoan2019a,Miranda-Villatoro2018,Forni2019}. 

For the plant in Section \ref{sec:fop}, we study the closed-loop behavior in the presence of nonlinearities and uncertainties by considering (nonlinear) perturbations on the sequestration rate $\eta$ and by replacing the linear feedback $\theta_1(u) = \theta_1 u $ with a nonlinear saturation such as the Hill function 
\begin{equation}
 \theta_1(u) = \frac{u^N}{k_1 + k_2 u ^N}.
\label{eq:Hill}
\end{equation}

From \eqref{eq:Hill}, the state-dependent open loop transfer function reads
\begin{equation}
\label{eq:robust_transfer_function} 
T_{\delta \bar{u} \to \delta \bar{y}}(z,s) = \frac{\theta_1'(z_1)\theta_2 \eta k z_1}{ s(\gamma + s)^2 (s + \eta (z_1+z_2))} ,
\end{equation}
which shows that the saturation essentially modulates the (linearized) feedback gain. $\theta_1$ is non-decreasing thus, from Theorem \ref{thm:rootlocus_nyquist}, the Nyquist plots in Figure \ref{fig:case1_0dom_d} remains compatible with $0$-dominance whenever $\max_{u\geq 0} \theta_1'(u) \leq 1$. Likewise, Figure \ref{fig:case1_2dom_f}, shows that compatibility with $2$-dominance should be guaranteed for any $\theta_1'(u) < \infty$.

\begin{figure}[t]
\begin{center}
\subfigure[$k_1 = 1, k_2 = 0.2, N = 1$]{\includegraphics[width=0.48\columnwidth]{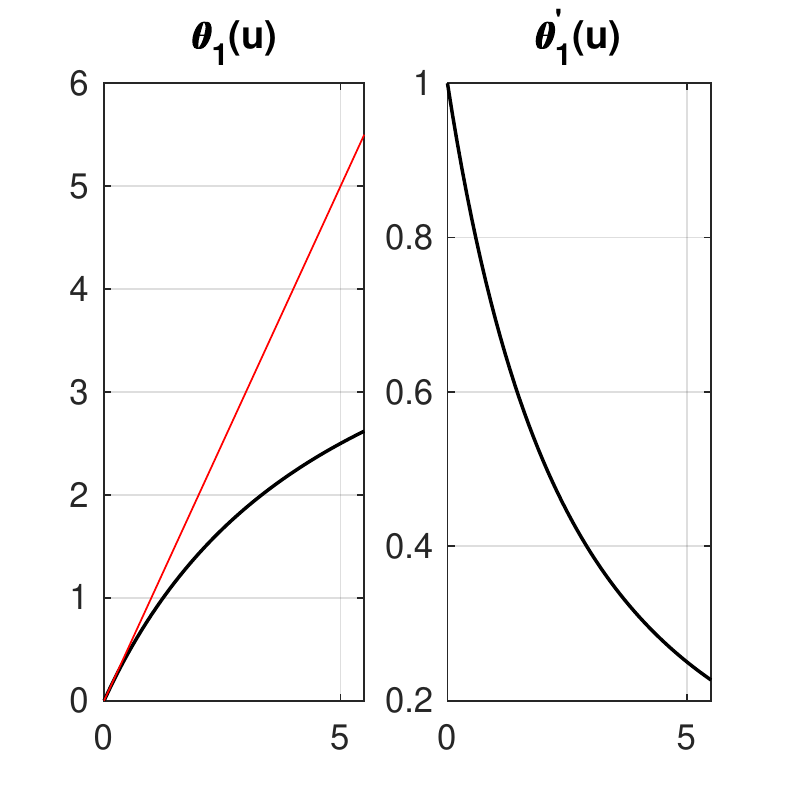} \label{fig:smallh}}    
\subfigure[$k_1 = 0.1, k_2 = 1, N = 2$]{\includegraphics[width=0.48\columnwidth]{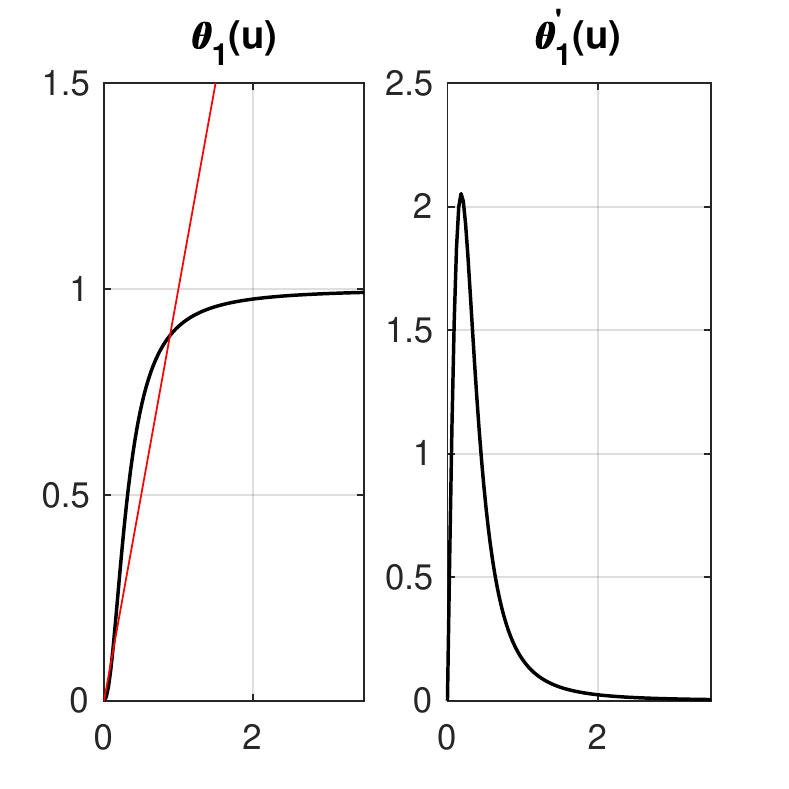} \label{fig:largeh}}
     \vspace{-2mm}
\subfigure[]{\includegraphics[width=0.48\columnwidth]{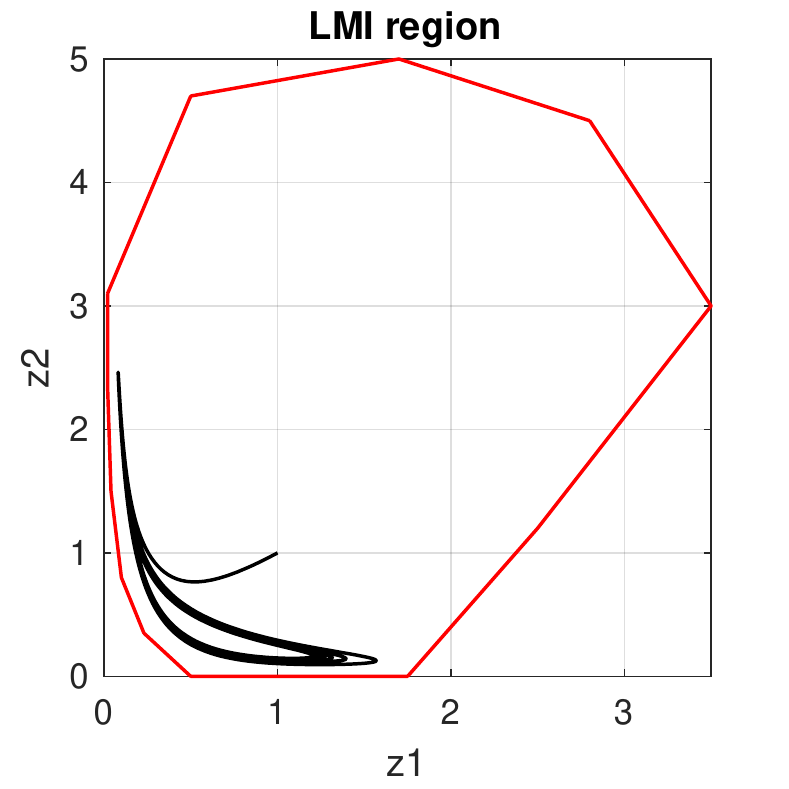} \label{fig:case1_2dom_smallh_c}}
\subfigure[]{\includegraphics[width=0.48\columnwidth]{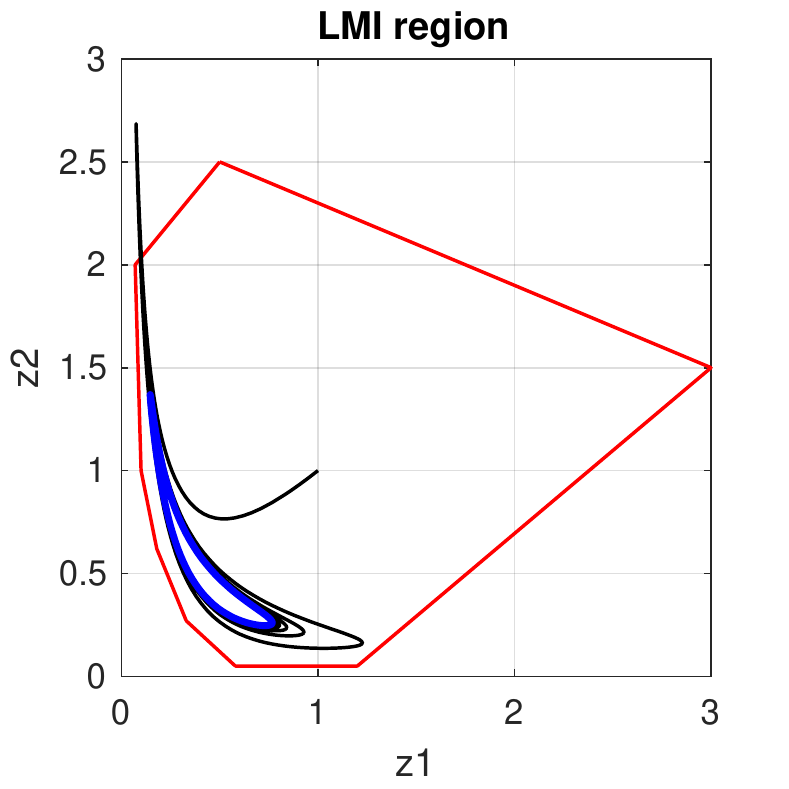} \label{fig:case1_2dom_largeh_d}}
\subfigure[]{\includegraphics[width=0.48\columnwidth]{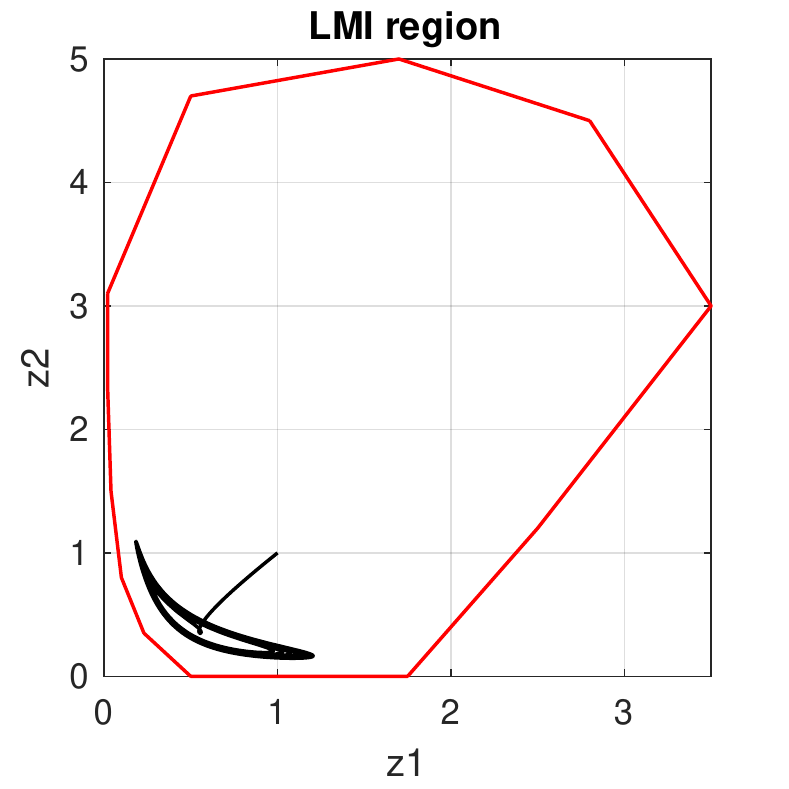} \label{fig:case1_2dom_smallh_e}} 
     \vspace{-2mm}  
\subfigure[]{\includegraphics[width=0.48\columnwidth]{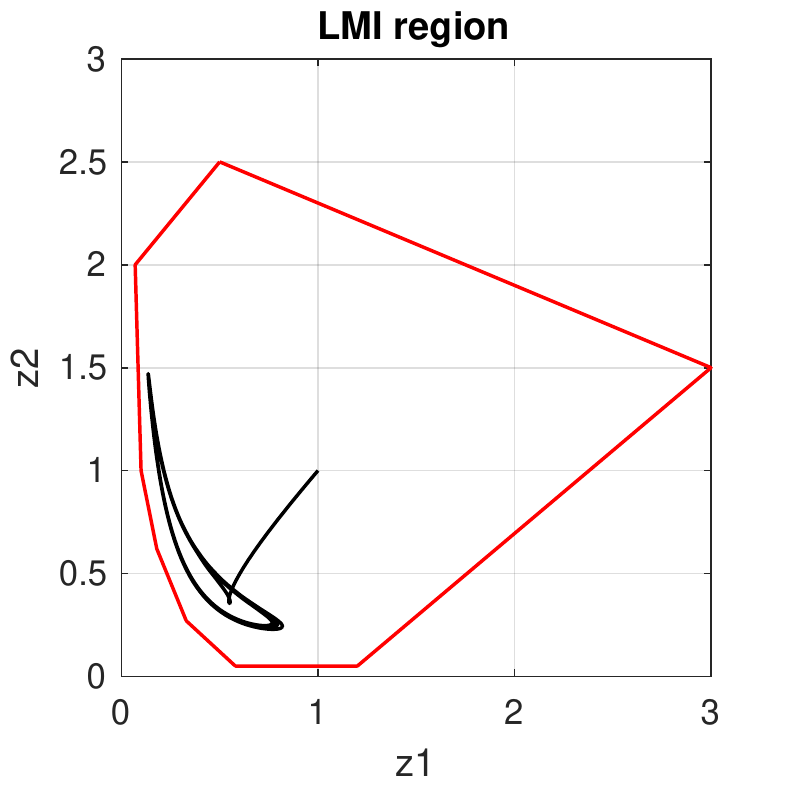} \label{fig:case1_2dom_largeh_f}}
\vspace{-1mm}
\caption{ \small
(a,b) -
    Hill functions.
(c,d,e,g) - trajectories projected on the $z$-plane. 
    $2$-dominance holds for $z$ constrained within the red curve
    and $x$ unconstrained. 
    Left column: simulations related to Hill function (a).
    Right column: simulations related to Hill function (b).
    (c,d) refer to $\theta_2=4$, $k=1$.
    (e,f) refer to $\theta_2=1$, $k=4$.
    The blue curve in (d) 
    shows the location of 
    the limit cycle projected on the $z$ axes,
    for clarity.
}
\label{fig:hill_functions_and_2dom}
\end{center}
\end{figure}

A complete analysis is beyond the scope
of the current paper. We look only at the two specific Hill 
functions in Figures \ref{fig:smallh} and 
\ref{fig:largeh}. The left one satisfies 
$\theta_1'(u) \in [0,1]$, thus it is compatible with $0$-dominance. 
The right one has a peak $\theta_1'(u) > 2$, which is compatible
with $2$-dominance but not with $0$-dominance.

The LMI \eqref{eq:robust_dominance_LMI}
certify the intuitive argument from the transfer function. 
For the Hill function in Figure \ref{fig:smallh}
and for $\theta_2=k=1$, 
$0$-dominance is preserved in a sizeable
region around the (new) fixed point. Likewise, 
$2$-dominance is preserved for both Hill functions, as shown
in Figure \ref{fig:hill_functions_and_2dom}. The figure
shows how the  nonlinearities
affect shape and position of the attractor but also
the region of feasibility of the LMI,  as illustrated by the
reduced red region in Figures \ref{fig:case1_2dom_largeh_d}
and \ref{fig:case1_2dom_largeh_f}.

\begin{figure}[t]
\begin{center}
\subfigure[]{\includegraphics[width=0.48\columnwidth]{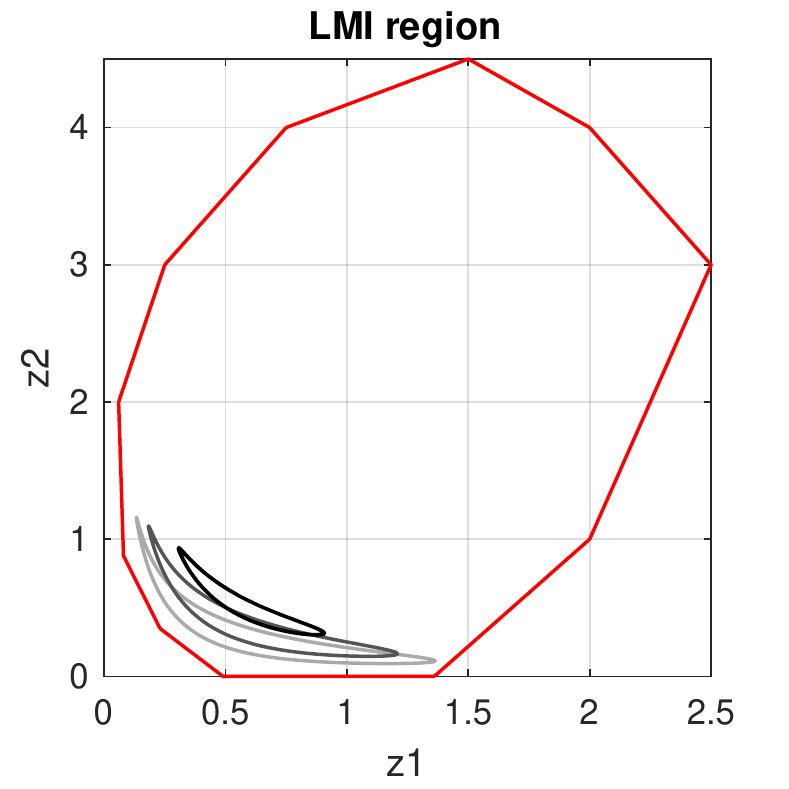} \label{fig:case1_2dom_smallh_eta_a}} 
     \vspace{-2mm}  
\subfigure[]{\includegraphics[width=0.48\columnwidth]{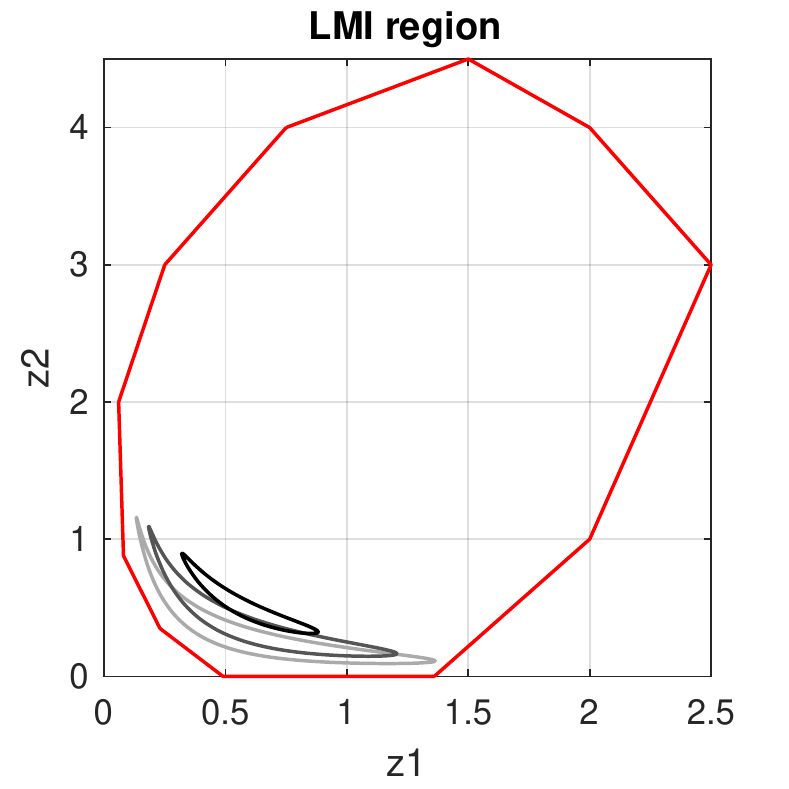} \label{fig:case1_2dom_smallh_eta_b}}
\vspace{-1mm}
\caption{ \small
(a) $\theta_2=4$, $k=1$;
(b) $\theta_2=1$, $k=4$.
Shaded gray attractors correspond to linear
binding rate $\eta \in \{7,10,13\}$ and Hill function activation in 
Figure \ref{fig:smallh}.
2-dominance  holds for $z$ constrained 
within the red curve and $x$ unconstrained.
} \label{fig:eta_and_2dom}
\end{center}
\end{figure}

To complete the analysis, we briefly look into
robustness to parametric uncertainties. For reasons
of space we consider only the closed loop based on 
the Hill function in Figure \ref{fig:smallh}. We replace the 
term $\eta z_1 z_2$ in \eqref{eq:sequestration}
with the uncertain monotonic nonlinear binding
\begin{equation}
    \eta(z_1, z_2): \qquad 0 < \ell_b \leq \partial \eta(z_1, z_2) \leq u_b 
    \label{eq:eta}
\end{equation}
From the transfer function \eqref{eq:robust_transfer_function} 
it is clear that the Nyquist diagram does not change significantly.
Indeed, 
for $\ell_b = 7$ and $u_b =13$ ($30\%$ variation on the nominal
value), the LMI \eqref{eq:robust_dominance_LMI}
certifies $2$-dominance for $z$ constrained within the red region 
of Figure \ref{fig:eta_and_2dom}.

We remark that achieving robust $2$-dominance
does not guarantee that oscillations persist, since
the attractor may reduce to an equilibrium.
A sufficient condition for the attractors within the region of 
$2$-dominance to be limit cycles is that every equilibrium 
in the region is unstable. This holds for the 
two Hill functions of Figure \ref{fig:hill_functions_and_2dom}.

\section{Future Directions}
There are a number of natural directions to extend this work. For the sake of clarity we focused so far on a constrained model class, however there is no fundamental barrier to analyzing other biological systems with similar control structure. For example, \cite{chen2012sequestration} showed that it is possible to use an AIF architecture along with positive feedback to produce bistability, such as in the model
\begin{equation*}
    \begin{split}
        \dot{z}_1 &= \mu_1 + \frac{\theta_1 z_1}{1 + \theta_1 z_1} - \eta z_1 z_2 -\gamma z_1 \\
        \dot{z}_2 &= \mu_2 - \eta z_1 z_2
    \end{split}
\end{equation*}
Where $\mu_1, \mu_2$ are transient inputs that can be used to switch the system from one state to the other. It is likely the case that such an architecture can be shown to exhibit $1$-dominance, and that even more complex plant models with the same positive feedback will still reduce to $1$-dominant dynamics.  

\begin{figure}[t]
\begin{center}
\vspace{1.5mm}
\includegraphics[width=.97\columnwidth]{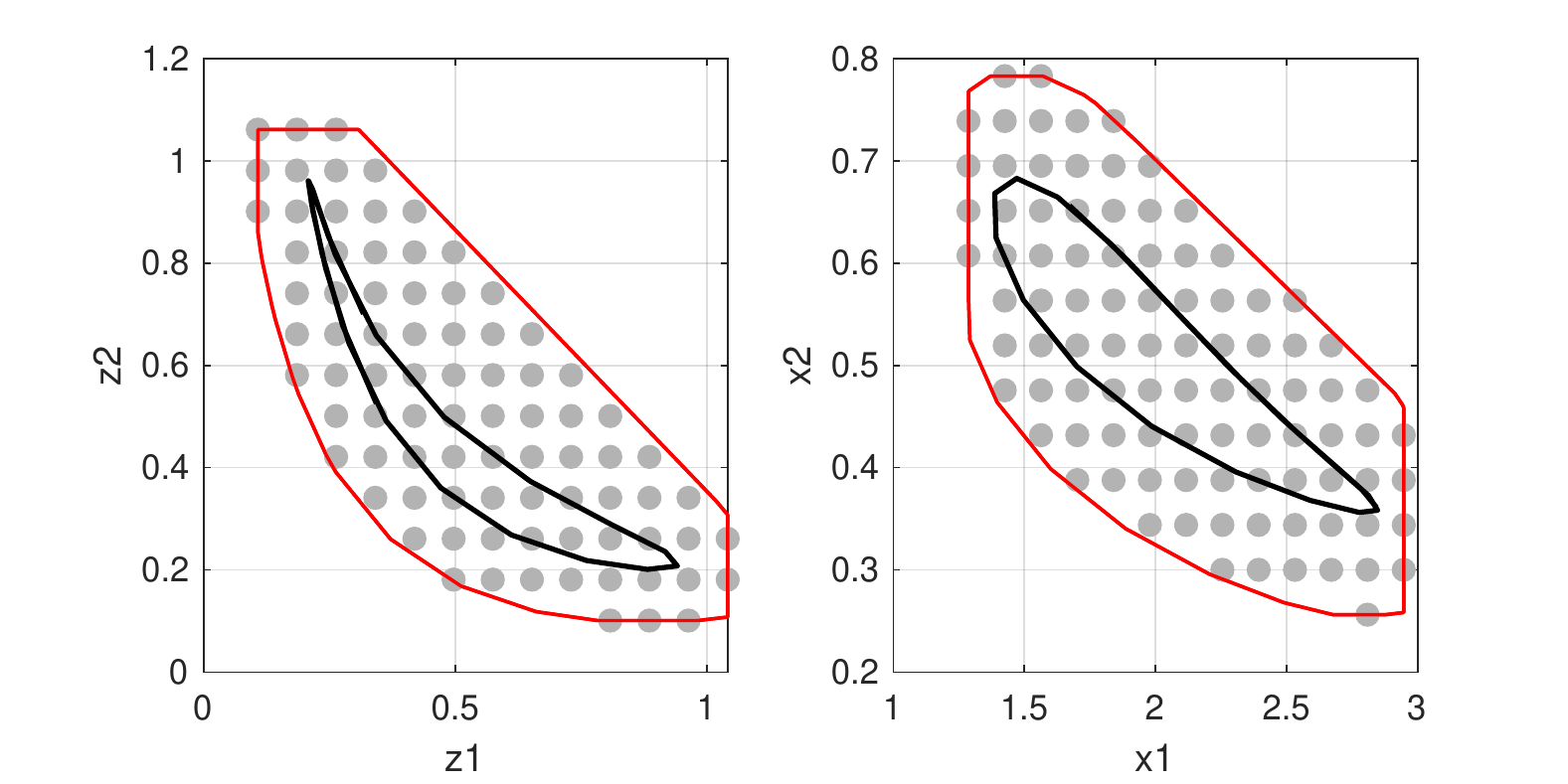} 
\vspace{-1mm}
\caption{ \small
Region of feasibility of the LMI \eqref{eq:dominance_LMI}
for the closed loop \eqref{eq:sequestration}, \eqref{eq:closed_loop_interconnection}, \eqref{eq:all_sequestration}. 
In contrast to our previous cases,
the nonlinear binding between $x_1$ and $x_2$ and between $x_1$ and $z_1$, respectively, leads to a region of feasibility that constraints both $z$ and $x$ species. 
Dominance is certified by the feasibility of a set of LMIs built
from the linearization at each shaded point.}
 \label{fig:all_sequestration}
\end{center}
\end{figure}

Alternatively, in \cite{olsman2019hard} a plant described by the dynamics 
 \begin{equation}
\Sigma_x:
\left\{
\begin{split}
    \dot{x_1} &= \phi_1 - \theta_1 x_1 u, \\
    \dot{x_2} &= \phi_2 - k x_1 x_2, \\
    y &= x_2
\end{split}
\right. 
\label{eq:all_sequestration}
\end{equation}
was shown via simulation to exhibit both a single locally stable equilibrium and a stable limit cycle for a particular set of parameters. It is likely possible to demonstrate this unusual form of bistability via dominance analysis. We would expect that such a system should not be described by global dominance behavior, but that it is possible to construct regions of $0$- and $2$-dominance that would characterize the
behavior of the system within those specific regions. Indeed, 
simulations show that the closed loop
\eqref{eq:sequestration}, \eqref{eq:closed_loop_interconnection}, \eqref{eq:all_sequestration} has stable oscillations for
$\mu = 2$, $\eta = 10$, $\theta_1 = k = \phi_1 = \phi_2 = 1$,
and $\theta_2 =4$, and our preliminary analysis 
shows that the closed loop
is $2$-dominant in the neighborhoods of its limit cycle,
as illustrated in Figure \ref{fig:all_sequestration}.

At technical level, we did not address how to build regions of LMI feasibility that contain the desired attractor / trajectories. This is fairly straightforward in practice, starting from simulations and building sufficiently large regions that safely contain the trajectories of interest. However, the analysis can be made more rigorous by  building (compact) regions of feasibility that are also forward invariant for the closed-loop dynamics. This additional property  would entail the stronger results that every trajectory starting in the region must converge to some (simple) attractor within the region.


\section{Conclusion}

The results presented here show how to use tools from the dominance theory to study nonlinear systems in synthetic biology. 
We analyzed a particular nonlinear circuit architecture, the antithetic integral feedback system, which is shown to be able to encode homeostatic regulation (0-dominance) and robust periodic oscillations (2-dominance).
For both cases, intuitive arguments based on the Nyquist criterion and root locus adapted to the linearized dynamics support parameter selection. Formal certificates are then provided by linear matrix inequalities. These certificates are inherently regional, in that they require the specification of a particular range of both state and parameter space.
Remarkably, the approach allows us to make statements about robustness for oscillatory regimes in much the same way we use classical robust control to analyze the robustness of equilibria.

Overall, our paper support two important ideas. First, that the AIF circuit should be thought of as a core component in synthetic biology
because of its capacity for diverse steady-state behaviors.
From this perspective, we might think the AIF circuit as the biological equivalent of an op-amp, playing a central role in enabling
monostable, multistable, and oscillatory circuits in synthetic biology.
Second, that these behaviors structurally arise from nonlinearity and require control tools that go beyond the stability analysis of a single equilibrium. Dominance theory makes useful steps in this direction. 

Our analysis is by no means comprehensive. We had made several simplifying assumptions and our analysis is limited to the specific set of parameters considered. However, we believe that the methodology described in this work presents progress towards a general approach to biological systems analysis. Dominance theory does not rely on the specific features of the nonlinearities, which makes it particularly well suited to biology, where models can be quite diverse and complex, yet the resulting dynamics are often surprisingly orderly and low-dimensional. \vspace{-2mm}

\begin{ack} \vspace{-3mm}
The authors would like to acknowledge the support of Johan Paulsson in the Systems Biology department at Harvard Medical School and Rodolphe Sepulchre in the Department of Engineering at the University of Cambridge.
\end{ack} \vspace{-5mm}

\bibliography{noah,fulvio}             
\end{document}